\documentclass[11pt,a4paper]{article}
\pdfoutput=1
\usepackage[]{inputenc,graphicx,amsmath,textcomp,amssymb}
\usepackage{subfigure}  
\newcommand{\be}{\begin{equation}}
\newcommand{\ee}{\end{equation}}
\newcommand{\bea}{\begin{eqnarray}}
\newcommand{\eea}{\end{eqnarray}}

\catcode`@=12

\begin{document}
\begin{flushright}
\end{flushright}
\thispagestyle{empty}
\begin{center}
{\Large\bf 
{Extension of Minimal Fermionic Dark Matter Model : \\ A Study
with Two Higgs Doublet Model }}\\
\vspace{1cm}
{{\bf Amit Dutta Banik} \footnote{email: amit.duttabanik@saha.ac.in}, 
{\bf Debasish Majumdar} \footnote{email: debasish.majumdar@saha.ac.in}}\\
\vspace{0.25cm}
{\normalsize \it Astroparticle Physics and Cosmology Division,}\\
{\normalsize \it Saha Institute of Nuclear Physics,} \\
{\normalsize \it 1/AF Bidhannagar, Kolkata 700064, India}\\
\vspace{1cm}
\end{center}
\begin{abstract}
We explore a fermionic dark matter model with a possible extension of Standard
Model (SM) of particle physics into two Higgs doublet model. Higgs doublets
couple to the singlet fermionic dark matter (FDM) through a non renormalisable
coupling providing a new physics scale. We explore the viability of such dark
matter candidate and constrain the model parameter space by collider serach,
relic density of DM, direct detection measurements of DM-nucleon scattreing
cross-section and with the experimentally obtained results from indirect search
of dark matter.     
\end{abstract}
\section{Introduction} 
The satellite borne experiments like Planck, WMAP etc. which 
study the anisotropies of cosmic microwave background radiations predict
that more than a quarter of the constituents of the Universe is  made of
unknown dark matter. The recent Planck data suggest that the relic abundance
for dark matter is within the range $\Omega_{\rm DM} {\rm h}^2 = 0.1199
\pm 0.0027$ \cite{Planck}, where $\rm h$ is the Hubble parameter normalised to 
100 km s$^{-1}$ Mpc$^{-1}$. There are also several ongoing terrestrial
experiments for direct detection of dark matter. Although no dark matter is
convincingly detected but there are claims of the observance of 
three potential dark matter signals by CDMS direct dark matter search
experiment \cite{CDMS,cdms15}. In addition
DAMA/NAI \cite{dama1}-\cite{dama3} dark matter direct search
experiment had also claimed to have observed the signature of the annual
modulation of dark matter signal $-$ a phenomenon that the dark matter direct
search signal should exhibit due to the revolution of earth around the Sun.
The ongoing direct search experiments such as XENON100 \cite{xenon} and
LUX \cite{lux}  give an upper bound in 
$m_{\chi}-\sigma_{\rm scat}$ plane where $\sigma_{\rm scat}$ is the dark 
matter elastic scattering cross-sections off the target detector and
$m_{\chi}$ is the dark matter mass. The XENON100 \cite{xenon} and LUX \cite{lux}
experiments provide stringet bounds on DM-nucleon scattering cross-section.
for different Dark matter masses. Dark matter particles can also be trapped
in a highly gravitating astrophysical objects and can eventually undergo 
annihilation to produce $\gamma$'s or fermion anti-fermion pairs. Such events
should show up as excesses over the expected abundance  of these particles in
the cosmos (for instance in cosmic rays). Indirect searches of dark matter by
detecting their annihilation products can be realised by looking for these
excesses in the Universe. In fact the satellite borne experiments such as Fermi-Lat
\cite{fermilat} (also known as Fermi gamma-ray space telescope or FGST),
Alpha Magnetic Spectrometer AMS \cite{ams} on board International Space Station
(ISS) or the earth bound experiments like H.E.S.S. \cite{hess}, MAGIC
\cite{magic} as also the Antarctica balloon-borne experiments like ATIC
\cite{atic} study the cosmic gamma ray, positron or antimatter excesses, origin
of which could be annihilation of dark matter at the regions of
astrophysical interest such as Galactic Centre (GC), Galactic halo etc. 

Although the dark matter (DM) searches are being vigorously persued, the 
particle constituent of dark matter is not known at all. Various  particle
physics models for cold dark matter  (CDM) are available in literature that
include the popular candidate neutralino which is supersymmetry motivated,
Kaluza Klein dark matter from theories of extra dimensions or particles in some
other proposed theories where simple extensions of Standard Model are
considered (such as adding a scalar singlet or an inert doublet and then
imposing a discrete ${\rm Z}_2$ symmetry that ensures the stability of the
dark matter candidate \cite{darkmatter}). In the present work we
consider an extension of Standard Model (SM) where a second Higgs doublet is
introduced in addition to the SM Higgs doublet. Though the recent findings
of CMS \cite{cms} and ATLAS \cite{atlas} have confirmed the existence of a SM
like scalar with mass  125 GeV, possibility of having a second Higgs doublet
accompanied by the SM sector Higgs doublet is not ruled out. Such an extension
of SM sector including a second Higgs doublet is preferably known as two Higgs
doublet model or THDM \cite{branco}. The two Higgs doublet model is the most 
general non supersymmetric extension of Standard Model (SM) when another 
complex doublet of same  hypercharge is added to the SM. Also a discrete 
symmetry is introduced between the Higgs doublets of THDM to avoid flavour
changing neutral current (FCNC) processes \cite{higgshunter}. 
In this work, we consider a singlet fermionic dark matter candidate
in THDM framework. Possibility of a singlet scalar appearing in THDM
to provide a feasible DM candidate has been studied extensively in
Refs. \cite{Aoki}-\cite{Martinez}.  
The case of low mass scalar DM in the framework of THDM has been presented
in a recent work by \cite{Drozd}. Thus, the dark matter candidate 
is the singlet fermion in our model. We then explore the viability
of this singlet fermion
for being a candidate of cold dark matter in the framework of THDM.
In a previous work \cite{kim}, a minimal model of singlet fermionic dark matter
is proposed which is formulated by adding a Lagrangian for the fermion to 
Standard Model Lagrangian. In the present work, however, we consider a THDM with an 
additional singlet fermion  which is treated as the DM candidate. Previous work
including fermionic dark matter in THDM  Ref. \cite{Cai}, is based on an ad-hoc 
assumption that the singlet dark matter couples to the SM Higgs ($h$) and does
not couple to the other scalar $H$ involved in THDM. Based on this simple
assumption the work by Cai. et. al. \cite{Cai} only explores the low mass dark matter
region ($m_{\rm DM}\leq20$ GeV). But in our case, the
singlet fermion, which is the DM candidate in the present model, couples to
both the Higgs doublets through a dimension five coupling when a new physics
scale $\Lambda$ is introduced. Hence, DM candidate in present scenario couples
to both the scalar bosons $h$ and $H$ of THDM. In addition the work by Cai et. al.
\cite{Cai} considered only type II THDM without exploring the THDM parameter space 
in the model and indirect detection of DM candidate is not taken into account.
However, in the present work both THDM phenomenology  and indirect DM detection
for type I and type II THDM are explored.
The stability of such a dark matter is ensured either by assigning a descrete
${\rm Z}_2'$ symmetry under which the singlet fermion is odd and the THDM 
sector is even or by assigning the baryon and lepton
charge of the singlet fermion to be zero as taken in Ref. \cite{kim}. In this
work we explore the possibility that within the framework of this model, the
fermion (added to the THDM) is a viable candidate for cold dark matter. We 
evaluate its direct detection cross-section and relic density and compare them
with the experimentally obtained results. The paper is organised as follows. 
In Sect.~\ref{S:model}, we introduce the model and describe the model 
parameters. The aspect of possible collider physics phenomenology for the model
is addressed in Sect.~\ref{S:LHC}. In Sect.~\ref{S:relic} we calculate the 
relic density of the  dark matter candidate in our proposed model. The model
parameters are constrained by comparing the calculated relic density with
observational dark matter relic density data obtained from PLANCK experiments.
Results for the allowed parameter space obtained from the relic density
calculation are presented in Sect.~\ref{S:results}. In Sect.~\ref{ss:dd}, we
calculate the spin independent direct detection scattering cross-section for
different masses of the present dark matter candidate. The model parameters are
then further constrained by results obtained from dark matter direct detection
experiments. Using the model parameter space thus constrained, we
study the indirect DM search for chosen benchmark points (BPs) in our model and 
compare them with the FGST (Fermi-LAT) results in Sect.~\ref{ss:id}. In 
Sect.~\ref{S:discussion} we summarise the work with concluding remarks and
discussions.
\section{The Model}
\label{S:model} 
In the present work we add a singlet fermion $\chi$ with two Higgs doublet
model. The singlet fermion $\chi$ in the resulting model, is the
dark matter candidate. The Lagrangian for $\chi$ can be written as
\bea
{\cal L}_{\chi} &=& \bar{\chi}i \gamma^{\mu}{\partial_{\mu}} 
\chi - m_0\bar{\chi}\chi \,\, .
\label{prob1}
\eea
As mentioned earlier, the stability of $\chi$ can be confirmed either by
assigning zero
lepton number and zero baryon number to the singlet fermion \cite{kim} or by
assuming a ${\rm Z}_2'$ symmetry under which $\chi$ is odd and the SM
sector is even. The total Lagrangian of the model in THDM framework can be
written as
\bea
{\cal L} &=& {\cal L}_{\rm THDM} + {\cal L}_{\chi} + {\cal L}_{\rm int}\,\, ,   
\label{prob2}
\eea
where ${\cal L}_{\rm int}$ denotes the interaction Lagrangian. The two Higgs
doublet model potential is expressed as
\bea
V(\Phi_1,\Phi_2) & = & m^2_1 \Phi^{\dagger}_1\Phi_1+m^2_2
\Phi^{\dagger}_2\Phi_2 + (m^2_{12} \Phi^{\dagger}_1\Phi_2+{\mathrm{h.c.}
}) +\frac{1}{2} \lambda_1 (\Phi^{\dagger}_1\Phi_1)^2 +\frac{1}{2}
\lambda_2 (\Phi^{\dagger}_2\Phi_2)^2\nonumber \\ 
& &+ \lambda_3
(\Phi^{\dagger}_1\Phi_1)(\Phi^{\dagger}_2\Phi_2) + \lambda_4
(\Phi^{\dagger}_1\Phi_2)(\Phi^{\dagger}_2\Phi_1) + \frac{1}{2}
\lambda_5[(\Phi^{\dagger}_1\Phi_2)^2+{\mathrm{h.c.}}] ~, 
\label{prob3}
\eea
where both the doublet Higgs fields $\Phi_1$ and $\Phi_2$ have non zero vacuum
expectation values and a discrete symmetry (${\rm Z}_2$) is imposed in between
the doublet fields in order to avoid FCNC processes. We consider a CP 
conserving two Higgs doublet model potential where all the parameters expressed
in Eq. \ref{prob3} are assumed to be real. In addition, the imposed discrete
symmetry ${\rm Z}_2$ will result in mainly four types of THDM namely type I,
type II, lepton specific and flipped THDM according to the nature of the 
coupling of fermions with the doublet fields. In the present work we consider
type I and type II THDM and construct the model. Thus the two scenarios we 
consider in this work are type I THDM + one singlet fermion and type II THDM
+ one singlet fermion.
Both the scenarios will give rise to  two charged Higgs fields
($H^{\pm}$), two CP even scalar fields ($h,H$), one CP odd scalar  ($A$) and
three Goldstone bosons ($G^{\pm},G$). The Higgs doublets $\Phi_1$ and $\Phi_2$
expressed in terms of physical states of the particles are written as
\cite{ferreira},
\bea
~~~~~~~~~~~~~~~~~\Phi_1 =\left( \begin{array}{c}
                           c_{\beta}G^+ - s_{\beta}H^+   \\
        \frac{1}{\sqrt{2}}(v_1+c_{\alpha}H-s_{\alpha}h+ic_{\beta}G-is_{\beta}A)  
                 \end{array}  \right)\,\, , 
\label{prob4}
\eea
\bea                   
~~~~~~~~~~~~~~~~~\Phi_2 =\left( \begin{array}{c}
                           s_{\beta}G^+ + c_{\beta}H^+   \\
        \frac{1}{\sqrt{2}}(v_2+s_{\alpha}H+c_{\alpha}h+is_{\beta}G+ic_{\beta}A)  
                 \end{array}  \right)\,\, ,
\label{prob5}
\eea
where $\tan{\beta} (= \frac{v_2}{v_1})$, is the ratio of the vacuum expectation
values $v_2$ and $v_1$ of the doublets $\Phi_1$ and $\Phi_2$ and $\alpha$ is 
the measure of mixing between two CP even scalars. The terms $c_x$ and $s_x$
($x = \alpha, \beta $) denote $\cos x$ and $\sin x$ respectively.
The scalar
potential for the THDM as expressed in Eq.~\ref{prob3} must be bounded from
below for the stability of vacuum. The Conditions for a stable vacuum for THDM are
\bea
\lambda_1,\,\lambda_2 > 0\, , ~~~~~~ 
\lambda_3 + 2\sqrt{\lambda_1\lambda_2}  >  0\, ,  ~~~~~~~
\lambda_3 +\lambda_4 -|\lambda_5| + 2\sqrt{\lambda_1\lambda_2}  >  0 \,\, .\nonumber
\eea   
Pertubative unitarity constraints for the THDM are also taken into account.
Bounds from the unitarity limits on THDM parameters are adopted from
\cite{branco}. 

The interaction Lagrangian, ${\cal L}_{\rm int}$ of dark matter fermion 
(Eq. \ref{prob2}) with $\Phi_1$ and $\Phi_2$ doublet fields is given by
\be
{\cal L}_{\rm int}=-\frac{g_1}{\Lambda}(\Phi_1^{\dagger}\Phi_1)\bar{\chi} \chi 
      -\frac{g_2}{\Lambda}(\Phi_2^{\dagger}\Phi_2)\bar{\chi} \chi\,\, , 
\label{prob6}
\ee
where $\Lambda $ is a high energy scale and $g_{1,2}$ are dimensionless 
couplings with the doublet fields $\Phi_{1,2}$. Interaction of THDM sector with
the DM candidate can now be obtained easily from Eqs. \ref{prob2}-\ref{prob6}. 
Dark matter fermion couples to both the physical Higgs particles $h$ and $H$
which are given by
\bea
g_{\bar{\chi}\chi h} = \frac{v}{\Lambda}(-g_1\sin\alpha \cos\beta +g_2\cos\alpha \sin\beta)\,\, , \nonumber \\
g_{\bar{\chi}\chi H} = \frac{v}{\Lambda}(g_2\cos\alpha \cos\beta +g_2\sin\alpha \sin\beta)\,\, ,
\label{prob7}
\eea
where $\Lambda$ being a very large scale with respect to $v$. Hence the 
couplings $g_{\bar{\chi}\chi h}$ and $g_{\bar{\chi}\chi H}$ are expected to be
small. Using Eqs.~\ref{prob1}-\ref{prob7}, mass of the singlet is expressed as
\bea
m_{\chi}=m_0 + v^2\left(\frac{g_1}{2\Lambda}\cos^2\alpha  + \frac{g_2}{2\Lambda}\sin^2\alpha \right)\,\, , \nonumber
\eea
where $v (= \sqrt{v_1^2 + v_2^2 })$, is 246 GeV.
Note that the new physics scale $\Lambda$ determines the coupling of DM 
particle to THDM sector and contributes significantly to the singlet 
fermion mass.
As mentioned earlier, the discrete ${\rm Z}_2$ symmetry imposed between the
Higgs doublets will result in four dfferent types of THDM. In this work we 
consider THDM of type I and type II. In type I THDM, only one scalar doublet 
(say $\Phi_2$) couples to the SM particles whereas in type II THDM, up type
quarks couple to one Higgs doublet and down type quarks and leptons couple to
the other. Higgs couplings to up type quarks, down type quarks and leptons in
case of type I THDM are given as \cite{higgshunter}
\bea
g_{\bar f f h}=-i\frac{gm_f}{2M_W}\frac{\cos\alpha}{\sin\beta}\,\, ,\hskip 15 pt
g_{\bar f f H}=-i\frac{gm_f}{2M_W}\frac{\sin\alpha}{\sin\beta}\,\, ,
\label{prob8}
\eea
where $f$ denotes all SM fermions (up quarks, down quarks and leptons) 
respectively. In case of type II THDM, Yukawa couplings are 
\bea
g_{\bar u u h}=-i\frac{gm_u}{2M_W}\frac{\cos\alpha}{\sin\beta}\,\, ,\hskip 15 pt
g_{\bar u u H}=-i\frac{gm_u}{2M_W}\frac{\sin\alpha}{\sin\beta}\,\, ,\nonumber \\
g_{\bar d d h}=-i\frac{gm_d}{2M_W}\frac{-\sin\alpha}{\cos\beta}\,\, ,\hskip 15 pt
g_{\bar d d H}=-i\frac{gm_d}{2M_W}\frac{\cos\alpha}{\cos\beta}\,\, ,\nonumber \\
g_{\bar l l h}=-i\frac{gm_l}{2M_W}\frac{-\sin\alpha}{\cos\beta}\,\, ,\hskip 15 pt
g_{\bar l l H}=-i\frac{gm_l}{2M_W}\frac{\cos\alpha}{\cos\beta}\,\, .
\label{prob9}
\eea
In the above, $u$ corresponds to up type quarks ($u,c,t$), $d$ correspondns to
down type quarks ($d,s,b$) and $l$ represents three families of leptons
($e,\mu, \tau$) respectively. Couplings to the gauge bosons ($V=W,Z$) for THDM
I and THDM II are same and given by \cite{higgshunter}
\bea
g_{W W h}=igM_W\sin(\beta-\alpha)g^{\mu\nu}\,\, ,\hskip 15 pt
g_{W W H}=igM_W\cos(\beta-\alpha)g^{\mu\nu}\,\, ,\nonumber \\
g_{Z Z h}=ig\frac{M_Z}{\cos\theta_W}\sin(\beta-\alpha)g^{\mu\nu}\,\, ,\hskip 15 pt 
g_{Z Z H}=ig\frac{M_Z}{\cos\theta_W}\cos(\beta-\alpha)g^{\mu\nu}\,\, .
\label{prob10}
\eea
In Eqs. \ref{prob8}-\ref{prob10}, $m_x$ ($x = u,~d,~l$ etc) represents the 
mass of quarks or leptons and $M_W$ and $M_Z$ denote the masses of $W$ and $Z$
bosons respectively. In the present framework with type I and type II THDM, we 
consider $h$ to be SM like Higgs boson with mass $m_h=125$ GeV and $H$ as the
non-SM Higgs with mass $m_H$.
\section{Collider physics phenomenology}
\label{S:LHC}
The existence of a scalar boson of mass 125 GeV has been confirmed by Large
Hadron Collider (LHC) \cite{cms,atlas}. In this work we treat the new found 
scalar boson to be equivalent to one of the CP even scalars ($h$) appearing in
THDMs. We further extend the model by including a possible fermionic dark matter
(FDM) candidate. This may necessarily affect the phenomenology of collider
physics. If the dark matter mass is small ($m_{\chi}\leq  m_{h}/2$) then one 
would expect an invisible decay of SM like Higgs boson ($h$) and the total 
decay width will change depending on the coupling constant 
$g_{\bar \chi \chi h}$ and other THDM parameters $\alpha,~\beta$. Since both
the scalar bosons in THDM couple with the DM fermion in the present framework,
it may change the standard bounds on THDM sector. The signal strength of SM 
like Higgs boson ($h$) to a specific channel for type I and type II THDM are 
given by
\bea
R_{I} = \frac{\sigma^{\rm I}_h}{\sigma^{\rm SM}} 
\frac{{\rm BR}^{\rm I}}{{\rm BR}^{\rm SM}}\,\, ,\hskip 15pt
R_{II} = \frac{\sigma^{\rm II}_h}{\sigma^{\rm SM}} 
\frac{{\rm BR}^{\rm II}}{{\rm BR}^{\rm SM}}\,\, , 
\label{prob11}
\eea 
where $\frac{\sigma^{\rm I,II}_h}{\sigma^{\rm SM}}$ represents the ratio of 
Higgs production cross-section in type I as also in type II THDM with respect to that
for SM ($\sigma^{\rm SM}$, is the SM Higgs production cross-scetion). The
branching ratio ($\rm BR$) to any specific channel
for the chosen model and for SM are given by ${{\rm BR}^{\rm X}}~,X=I,II$ and 
${{\rm BR}^{\rm SM}}$. The ratio $\frac{\sigma^{\rm X}_h}{\sigma^{\rm SM}}$ 
($X=I,II$) in Eq.~\ref{prob11} for 125 GeV Higgs boson can be expressed as
\be
\frac{\sigma^X_h}{\sigma^{\rm SM}}=\frac{\sigma_{tt}f^2_t+\sigma_{bb}f^2_b+\sigma_{tb}f_t f_b}{\sigma^{\rm SM}}\,\, ,
\label{prob12}
\ee
where $\sigma_{tt}$, $\sigma_{bb}$ are the Higgs production cross-sections
from top and bottom quarks respectively and $\sigma_{tb}$ is the contribution
from top-bottom interference. For the calculation of SM Higgs signal strength,
we have adopted the leading order (LO) production cross-sections obtained 
from \cite{1312.5571v2}. The factors $f_t,~f_b$ in Eq.~\ref{prob11} are the 
Yukawa couplings of SM like Higgs ($h$) with top and bottom quarks for the 
specific model nomalised with respect to SM. For type I THDM,
$f_t=f_b=\frac{\cos \alpha}{\sin \beta}$ and for type II THDM these factors are
given as $f_t= \frac{\cos \alpha}{\sin \beta}$ and  
$f_b= \frac{-\sin \alpha}{\cos \beta}$. As defined earlier, $\alpha$ is the
mixing angle between the CP even scalars $h$ and $H$ and $\beta$ is given by
the ratio of the VEVs $v_2$ and $v_1$ of Higgs doublets $\Phi_2$ and $\Phi_1$
respectively ($\tan\beta=\frac{v_2}{v_1}$). ATLAS and CMS experiments have
measured the signal strengths of SM Higgs ($h$) boson to different production
channels such as $b\bar b, \tau \bar{\tau},\gamma\gamma,WW^*,ZZ^*$. The mean
signal strengths of SM Higgs to these channels measured by ATLAS and the best
fit value of combined signal strength of $h$ given by CMS experiment are found
to be \cite{atlas1,cms1}
\be
R_{\rm ATLAS} = 1.23 \pm{0.18}\,\, , \hskip 15 pt  
R_{\rm CMS} = 0.8 \pm{0.14}\,\, . 
\label{prob13}
\ee
In the present scenario with THDM, we have a non-SM Higgs ($H$) in addition to
the SM scalar $h$. The signal strengths of non-SM Higgs boson for type I and
type II THDM are given as  
\bea
R'_{I} = \frac{\sigma^{\rm I}_H}{\sigma'^{\rm SM}} 
\frac{{\rm BR'}^{\rm I}}{{\rm BR'}^{\rm SM}} \,\, {\rm and} \,\, 
R'_{II} = \frac{\sigma^{\rm II}_H}{\sigma'^{\rm SM}} 
\frac{{\rm BR'}^{\rm II}}{{\rm BR'}^{\rm SM}}\,\,  
\label{prob14}
\eea
respectively, where $\sigma^{\rm X}_H$ ($\rm{ X=I,~II}$ depending on the nature
of THDM considered) is the non-SM Higgs production cross-section and
${\rm BR'}^X$ is the branching ratio of $H$ to any specific channel. In 
Eq.~\ref{prob14}, $\sigma'^{\rm SM}$ and ${\rm BR'}^{\rm SM}$ represent the
production cross-section and branching ratio of the non-SM Higgs boson ($H$)
with mass $m_H$. The modified non-SM Higgs production cross-section ratio can
be given as
\be
\frac{\sigma^X_H}{\sigma'^{\rm SM}}=\frac{\sigma'_{tt}f'^2_t+\sigma'_{bb}f'^2_b+\sigma'_{tb}f'_t f'_b}
{\sigma'^{\rm SM}}\,\,.
\label{prob15}
\ee
Similar to Eq.~\ref{prob11}, in Eq.~\ref{prob15} also, the factors $f'_t,~f'_b$
are the  SM normasiled Yukawa 
couplings of non-SM Higgs $H$ with top and bottom quarks. For the case of type
I THDM, $f'_t=f'_b=\frac{\sin \alpha}{\sin \beta}$, whereas those for type II 
THDM are $f'_t=\frac{\sin \alpha}{\sin \beta}$ and $f'_b=\frac{\cos \alpha}{\cos \beta}$.
In the present work we consider two values of non-SM Higgs mass and they are
chosen as $m_H=$ 150 GeV and 200 GeV. The calculations are performed for each 
of these chosen masses.
We use the leading order production cross-section 
($\sigma'_{tt},~\sigma'_{bb},~\sigma'_{tb}$ and $\sigma'^{SM}$) obtained from 
Ref. \cite{1312.5571v2} for the chosen $m_H$ values in the work. 
Invisible decay of the non-SM Higgs (for
$m_\chi \leq m_H/2$) has also been taken into account. Since no signature of
additional Higgs has been reported by ATLAS and CMS experiment, it is likely to
assume that the non-SM Higgs signal strength is negligibly small compared to
that of SM Higgs. Hence, throughout the work, we restrict the signal strength
for non-SM scalar satisfying the condition $R'_X\leq 0.2~(X=I,II)$. SM branching
ratios for specifc decay modes of SM Higgs (${\rm BR}^{\rm SM}$ with mass 
$m_h=125$ GeV) and non-SM Higgs (${\rm BR'}^{\rm SM}$ for $m_H=150$ and 200 GeV)
are adopted from Ref.~\cite{Denner}. It is to be mentioned that in this work we do 
not consider any ad-hoc condition, e.g. by setting 
$g_{\bar{\chi}\chi h}=0$ or $g_{\bar{\chi}\chi H} =0$
\cite{Cai} for the SM like scalar (assuming $\sin(\beta-\alpha)=\pm1$
when $h$ is SM like or $\sin(\beta-\alpha)=0$ when $H$ is SM like).
In the present formalism we consider the total allowed range of availbale parameter space
independent of these conditions and restrict them by using limits on SM Higgs signal strength 
from CMS and ATLAS (Eq.~\ref{prob13}).  
\section{DM annihilation and relic density}
\label{S:relic}
In order to evaluate the relic density of the fermionic dark matter candidate
proposed in this work one requires to solve the Boltzmann equation \cite{kolb}
\bea
\frac{d n}{d t} + 3 {\rm H}n &=& - \langle \sigma v \rangle (n^{2}-n_{eq}^{2})\,\, 
\label{prob16}
\eea
where $n$ is the actual number density of the particle species, 
$\rm H$ is the Hubble parameter and $n_{eq}$ is the number density at 
thermal equilibrium. An approximate expression for relic density 
$\Omega$ or $\Omega {\rm h}^2$ 
($\rm h = \rm H/(100\, {\rm km} {\rm s}^{-1} {\rm Mpc}^{-1})$) 
that can be obtained from Eq. \ref{prob16} is given by   
\bea
\Omega_{\rm DM}{\rm h}^2 &=& \frac{1.07\times 10^9 x_F}
{\sqrt{g_*}M_{\rm Pl}\langle \sigma v\rangle}
\label{prob17}
\eea
where $x_F=m_{\chi}/T_F$, $g_*$ is the effective degrees of freedom and
$M_{\rm Pl}=1.22\times10^{19}$ is the Planck mass.
The particle physics input to Eqs. \ref{prob16}-\ref{prob17}
is the thermal averaged annihilation cross-section  
$\langle \sigma v \rangle$ and one needs to calculate this quantity for the
present fermionic dark matter candidate in our model. The freeze out
temperature $T_F$ (or $x_F$) in Eq.~\ref{prob17} can be computed by iteratively
solving the equation
\bea
x_F &=& \ln \left ( \frac{m_{\chi}}{2\pi^3}\sqrt{\frac{45M_{Pl}^2}{2g_*x_F}}
\langle \sigma v \rangle \right )\,\, . 
\label{prob18}
\eea
The freeze out temperature thus obtained is then used to evaluate the relic
density of the dark matter candidate $\chi$ in our model. In order to solve
for the freeze out temperature, it is therefore essential to calculate the
annihilation cross-section of the dark matter candidate.  Dark matter 
candidates in the present model annihilate to SM particles through $h$ or $H$
mediated s-channel proceses. The total annihilation cross-section  $\sigma v$
can be expressed as a sum of the three terms
\bea
\sigma v &=& (s-4m_{\chi}^2) \left [ A\frac{1}{(s-m_h^2)^2+m_h^2\Gamma_h^2}+
B\frac{1}{(s-m_H^2)^2+m_H^2\Gamma_H^2} \right . \nonumber \\ 
&& \left . +C\frac{2(s-m_h^2)(s-m_H^2)+2m_hm_H\Gamma_h\Gamma_H}
{[(s-m_h^2)^2+m_h^2\Gamma_h^2][(s-m_H^2)^2+m_H^2\Gamma_H^2]} \right ]\, .
\label{prob19}
\eea
In Eq. \ref{prob19}, $\Gamma_h$ and $\Gamma_H$ are decay widths of light Higgs
($h$) and heavy Higgs particle ($H$) respectively. We set the light Higgs mass
$m_h$ to be 125 GeV and consider each of the two values of non-SM Higgs mass 
$m_H=$ 150 GeV and 200 GeV. Thus we assume $m_H > m_h$ in the present work. 
The terms $A,~B$ and $C$ in the expression for $\sigma v$ (Eq. \ref{prob19})
in case of THDM I are given as (with summation convention imposed on quarks and
leptons)
\bea
A &=& g_{\bar \chi \chi h}^2\frac{G_F}{4\pi\sqrt{2}}
\left [\frac{c_{\alpha}^2}{s_{\beta}^2}(N_cm_{u_i}^2\gamma_{u_i}^3
+N_cm_{d_i}^2\gamma_{d_i}^3+
m_{l_i}^2\gamma_{l_i}^3) \right . \nonumber \\
&& \left . +\frac{1}{2}s_{\beta-\alpha}^2s(1-x_W+\frac{3}{4}x_W^2)\gamma_W+
\frac{1}{4}s_{\beta-\alpha}^2s(1-x_Z+\frac{3}{4}x_Z^2)\gamma_Z \right ]\,,
\label{prob20}
\eea
\bea
B &=& g_{\bar \chi \chi H}^2\frac{G_F}{4\pi\sqrt{2}}
\left [\frac{s_{\alpha}^2}{s_{\beta}^2}(N_cm_{u_i}^2\gamma_{u_i}^3
+N_cm_{d_i}^2\gamma_{d_i}^3+
m_{l_i}^2\gamma_{l_i}^3) \right . \nonumber \\
&& + \left . \frac{1}{2}c_{\beta-\alpha}^2s(1-x_W+\frac{3}{4}x_W^2)\gamma_W+
\frac{1}{4}c_{\beta-\alpha}^2s(1-x_Z+\frac{3}{4}x_Z^2)\gamma_Z \right ]\,,
\label{prob21}
\eea
and
\bea
C&=&g_{\bar \chi \chi h}g_{\bar\chi \chi H}\frac{G_F}{4\pi\sqrt{2}}
\left [\frac{c_{\alpha}s_{\alpha}}{s_{\beta}^2}(N_cm_{u_i}^2\gamma_{u_i}^3
+N_cm_{d_i}^2\gamma_{d_i}^3 +
m_{l_i}^2\gamma_{l_i}^3) \right. \nonumber \\
&& +\left . \frac{1}{2}s_{\beta-\alpha}c_{\beta-\alpha}s(1-x_W+\frac{3}{4}x_W^2)\gamma_W+
\frac{1}{4}s_{\beta-\alpha}c_{\beta-\alpha}s(1-x_Z+\frac{3}{4}x_Z^2)\gamma_Z \right ]\,.
\label{prob22}
\eea
For type II THDM, the expressions for $A,~B$ and $C$ are
\bea
A &=& g_{\bar \chi \chi h}^2\frac{G_F}{4\pi\sqrt{2}}
\left [N_cm_{u_i}^2\frac{c_{\alpha}^2}{s_{\beta}^2}\gamma_{u_i}^3
+N_cm_{d_i}^2\frac{s_{\alpha}^2}{c_{\beta}^2}\gamma_{d_i}^3+
m_{l_i}^2\frac{s_{\alpha}^2}{c_{\beta}^2}\gamma_{l_i}^3 \right .\nonumber \\
&& + \left. \frac{1}{2}s_{\beta-\alpha}^2s(1-x_W+\frac{3}{4}x_W^2)\gamma_W+
\frac{1}{4}s_{\beta-\alpha}^2s(1-x_Z+\frac{3}{4}x_Z^2)\gamma_Z \right ]\,,
\label{prob23}
\eea
\bea
B &=& g_{\bar \chi \chi H}^2\frac{G_F}{4\pi\sqrt{2}}
\left [N_cm_{u_i}^2\frac{s_{\alpha}^2}{s_{\beta}^2}\gamma_{u_i}^3
+N_cm_{d_i}^2\frac{c_{\alpha}^2}{c_{\beta}^2}\gamma_{d_i}^3+
m_{l_i}^2\frac{c_{\alpha}^2}{c_{\beta}^2}\gamma_{l_i}^3 \right . \nonumber \\
&& + \left . \frac{1}{2}c_{\beta-\alpha}^2s(1-x_W+\frac{3}{4}x_W^2)\gamma_W+
\frac{1}{4}c_{\beta-\alpha}^2s(1-x_Z+\frac{3}{4}x_Z^2)\gamma_Z\right ]\,,  
\label{prob24}
\eea
\bea
C &=& g_{\bar \chi \chi h}g_{\bar \chi \chi H}\frac{G_F}{4\pi\sqrt{2}}
\left [N_cm_{u_i}^2\frac{s_{\alpha}}{s_{\beta}}\frac{c_{\alpha}}{s_{\beta}}\gamma_{u_i}^3
-N_cm_{d_i}^2\frac{c_{\alpha}}{c_{\beta}}\frac{s_{\alpha}}{c_{\beta}}\gamma_{d_i}^3-
m_{l_i}^2\frac{c_{\alpha}}{c_{\beta}}\frac{s_{\alpha}}{c_{\beta}}
\gamma_{l_i}^3  \right .\nonumber \\
&& + \left . \frac{1}{2}c_{\beta-\alpha}s_{\beta-\alpha}s(1-x_W+\frac{3}{4}x_W^2)\gamma_W+
\frac{1}{4}c_{\beta-\alpha}s_{\beta-\alpha}s(1-x_Z+\frac{3}{4}x_Z^2)\gamma_Z \right ]\,.   
\label{prob25}
\eea
In all the above expressions (Eqs.~\ref{prob20}-\ref{prob25}) 
$\gamma_a= (1-\frac{4m_a^2}{s})^{\frac{1}{2}}$ ($a = u,~d,~l,~W,~Z$), 
$X_B=\frac{4m_B^2}{s}$ and $N_c = 3$ for quarks.
Thermal average of pair annihilation cross-section of DM to SM particles is 
given by
\bea
\langle \sigma v \rangle=\frac{1}{8m_{\chi}^4T_FK_2^2(m_{\chi}/T_F)}\int_{4m_{\chi}^2}^\infty ds~\sigma (s)~(s-4m_{\chi}^2) \sqrt{s}
K_1(\sqrt{s}/T_F) ,    
\label{prob26}
\eea
where $K_1$ and $K_2$ are modified Bessel function. Using Eqs. 
\ref{prob19}-\ref{prob26}, the annihilation cross-section 
$\langle \sigma v \rangle$  of DM candidate into SM particles is evaluated for
both type I and type II THDM. We first solve for the freeze out temperature
$T_F$ using Eq.~\ref{prob18}. The relic  density $\Omega_{\rm DM}{\rm h}^2$ of
dark matter is obtained by solving Eq.~\ref{prob17} 
in order to satisfy dark matter relic density obtained from PLANCK experimental 
value $\Omega_{\rm DM}{\rm h}^2=0.1199\pm0.0027$\cite {Planck}. The DM relic
density is computed with the chosen model parameters
such that \footnote{We have checked that in order to satisfy the PLANCK
results, these ranges of the parameters suffice.}
\bea
m_{\chi}\leq 300~{\rm GeV}\,\, , \nonumber \\
10^{-4}\leq |g_{\bar{\chi}\chi h}| \leq 0.1\,\, , \nonumber \\
10^{-4}\leq |g_{\bar{\chi}\chi H}| \leq 0.1\,\, , \nonumber \\
-\pi/2\leq \alpha \leq \pi/2\,\, , \nonumber \\
1\leq \tan \beta < 30\,\, .
\label{prob27}
\eea 
As mentioned earlier, the calculation of dark matter relic density is performed
for two values of non-SM scalar mass $m_H$ taken to be 150 GeV and 200 GeV.
We further constrain the model 
parameter space using the bounds for SM Higgs signal strength as obtained from
ATLAS and CMS experiments (Section~\ref{S:LHC}) as also using the
bounds on the signal strength of $H$ ($R'_{I,II}\leq 0.2$).
\section{Results}
\label{S:results}
\begin{figure}[h!]
\centering
\subfigure[]{
\includegraphics[height=5 cm, width=5 cm,angle=0]{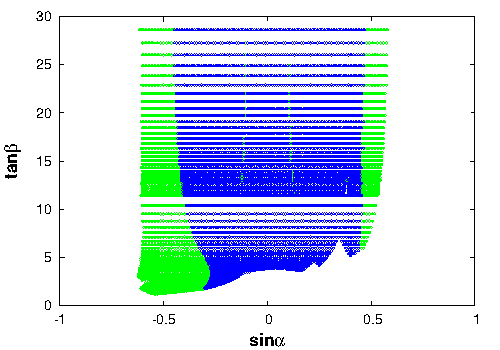}}
\subfigure[]{
\includegraphics[height=5 cm, width=5 cm,angle=0]{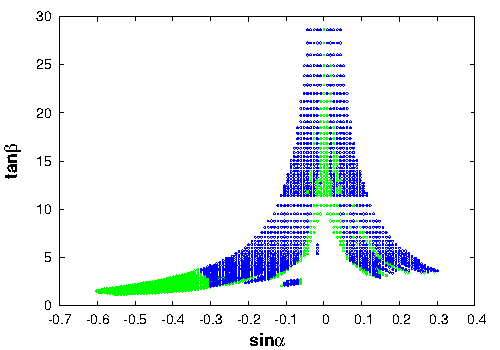}}
\subfigure []{
\includegraphics[height=5 cm, width=5 cm,angle=0]{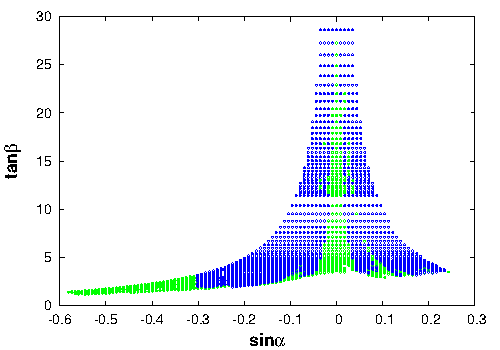}}
\caption{Allowed $\sin\alpha$-$\tan \beta$ parameter space for type I
THDM consistent with $R_{\rm CMS}$ within the framework of present DM model
(Fig.~\ref{fig1}a). Green and blue coloured regions are for $m_H=$ 150 geV and
200 GeV respectively. Similar plots for the case of type II THDM is shown in
Fig.~\ref{fig1}b. Valid parameter space in typr II THDM satifying 
$R_{\rm ATLAS}$  only is depicted in Fig.~\ref{fig1}c. The computation for all
the plots are performed with the constrained range of model parameter space 
values which produce required DM relic density consistent with PLANCK results.
For all the plots the constrained $R'_{I,II}$ is respected.}
\label{fig1}
\end{figure}
In this section we present the results for our fermionic dark matter in type I
and type II THDM. We first obtain the relic density oid DM candidate by solving
the Boltzmann equation (Eq.~\ref{prob16}). The Boltzmann equation is solved by 
using the range of parameter space given in Eq.~\ref{prob27} and the relic
density of the fermionic dark matter in the present model is then calculated.
The comparison with the PLANCK's result for DM relic density, constraints the
parameter space of the model considered in this work. The signal strength $R_X$
($X=I,~II$; $I,~II$ corresponds to type I and type II THDM respectively) for
the SM Higgs $h$ is computed with the parameter space restricted by PLANCK
results. As mentioned earlier, we also compute the signal strength
$R'_X$, the signal strength of the other Higgs $H$ and its value is kept in
the limit $R'_X\leq 0.2$. The calculated values of both $R_I$ and $R_{II}$ are
compared with the CMS and ATLAS limits for the SM signal strength. Thus the 
parameter space is further constrained by the CMS and ATLAS results.   
In Fig.~\ref{fig1}a-c  we show the allowed parameter space in
$\sin \alpha$-$\tan \beta$ plane for fermionic dark matter for each of
type I and type II THDM scenarios extended with FDM. The plots in 
Fig.~\ref{fig1} are obtained for two 
values of $H$ mass namely $m_H=$ 150 and 200 GeV.  In Fig.~\ref{fig1}a the 
variations of $\sin \alpha$ with $\tan \beta$ for FDM extended type
I THDM are shown.We found that for type I THDM along with FDM fails to satisfy
the combined signal strength as predicted by ATLAS ($R_{\rm ATLAS}$). Hence
in Fig.~\ref{fig1}a, only the constraints from CMS experimental results (for 
signal strength, i.e. $R_{\rm CMS}$) are imposed.
The blue and green scattered regions in Fig.~\ref{fig1}a-\ref{fig1}c
represent the respective allowed parameter space when $m_H$ is chosen to be
150 GeV and 200 GeV repectively.
It can also be observed from Fig.~\ref{fig1}a that increase in
the mass of the other scalar $H$ associated with the model results in 
considerable reduction in the overall allowed THDM parameter space. In
Fig.~\ref{fig1}b we plot the available region of 
$\sin \alpha$- $\tan \beta$ plane for the case of fermionic dark 
matter in type II THDM consistent with the PLANCK relic density as als0 SM
Higgs signal strength $R_{\rm CMS}$ given in Eq.~\ref{prob13} with 
$R'_{II}\leq 0.2$.
Similar allowed regions but $R_{\rm CMS}$ replaced with $R_{\rm ATLAS}$ 
(ATLAS bound) are shown in Fig.~\ref{fig1}c for type II THDM scenario.
For type II THDM, we use the same colour convention as used in the case of type
I THDM (Fig.~\ref{fig1}a) to show the valid region of parameter space for
$m_H=$150 and 200 GeV. Comparison of the plots in  Fig.~\ref{fig1}b-c with the
type I THDM case (Fig.~\ref{fig1}a) clearly shows that there is less allowed
parameter space available for type II THDM. It is to be noted that for type II
THDM involving FDM is in agreement with both the combined (for all five 
channels namely $b\bar b, \tau \bar{\tau},\gamma\gamma,WW^*,ZZ^*$) signal
strengths $R_{\rm CMS}$ and $R_{\rm ATLAS}$ as predicted independently by CMS
and ATLAS experimental results. Note that, for the case of type II THDM shown
in Fig.~\ref{fig1}b-c too, the avialable region of
$\sin \alpha$-$\tan\beta$ plane decreases with increase of the mass of
$H$ which is similar to the trend observed for type I THDM formalism 
(Fig.~\ref{fig1}a). 
\subsection{Direct detection measurements}
\label{ss:dd}
\begin{figure}[h!]
\centering
\subfigure[]{
\includegraphics[height=5 cm, width=5 cm,angle=0]{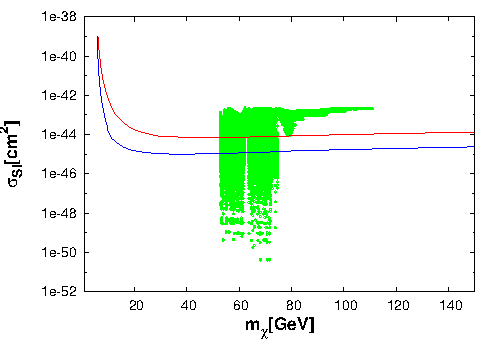}}
\subfigure []{
\includegraphics[height=5 cm, width=5 cm,angle=0]{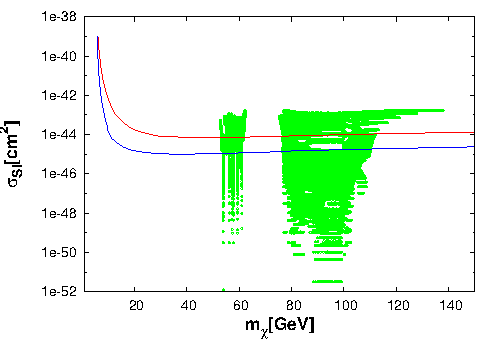}}
\subfigure[]{
\includegraphics[height=5 cm, width=5 cm,angle=0]{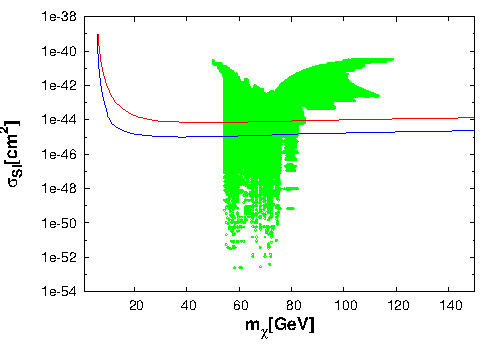}}
\subfigure []{
\includegraphics[height=5 cm, width=5 cm,angle=0]{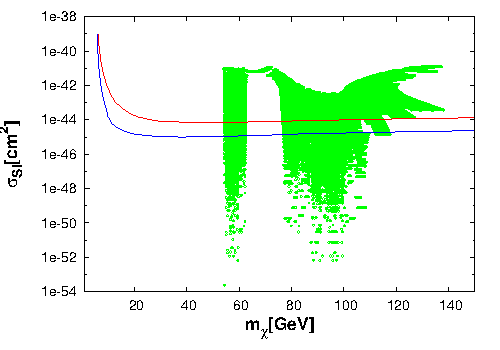}}
\subfigure[]{
\includegraphics[height=5 cm, width=5 cm,angle=0]{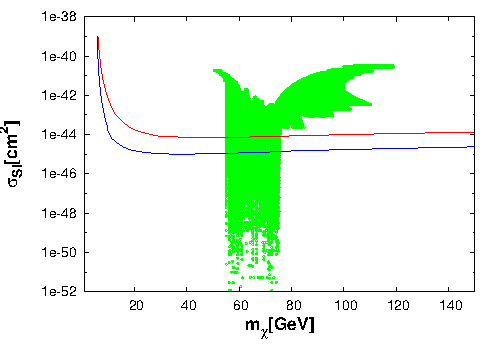}}
\subfigure []{
\includegraphics[height=5 cm, width=5 cm,angle=0]{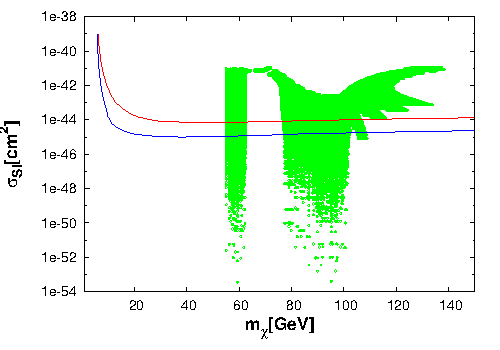}}
\caption{Fig.~\ref{fig2}a-b shows the $m_{\chi}-\sigma_{\rm SI}$ parameter
space for FDM in type I THDM iallowed by PLANCK relic density and collider
bounds plotted using $R_{\rm CMS}$ for $m_{H}=$
150 and 200 GeV. Similar plots in $m_{\chi}-\sigma_{\rm SI}$ plane with type
II THDM are shown in Fig.~\ref{fig2}c-d whereas the plots in Fig.~\ref{fig2}e-f
are in agreement with $R_{\rm ATLAS}$. All the results in Fig.~\ref{fig2}a-f 
also satisfy the bounds from DM relic density and $R'_{I,II}\leq$ 0.2. The
red and blue lines are respective bounds on DM-nucleon scattering cross-section
from XENON100 and LUX DM direct search experiments.}  
\label{fig2}
\end{figure}
We further restrict the allowed parameter space of our model with the
direct detection experimental bounds on DM-nucleon scattering cross-section.
Direct detection of dark matter utilises the phenomenon of a possible elastic
scattering off a nucleus of detecting material. In order to enable a uniform
comparison of experimental results from different dark matter experiments with
diffrent detecting materials, the experimentally obtained DM-nucleus elastic
scattering cross-section ($\sigma_{\rm scat}$) is reduced to DM-nucleon 
scattering cross-section. The experimental results are then expressed as the allowed
region in $m_{\chi}-\sigma_{\rm scat}^{\rm nucleon}$ plane. This elastic
scattering cross-section can be spin independent (SI) or spin dependent (SD),
depending on the ground state spin of detector nucleus. The elastic scattering
of the dark matter particle off the target causes the recoil of the target
nucleus. This recoil energy is measured in the experiment and allowed region in
the plane of scattering  cross-section and dark matter mass is then obtained.
The spin independent dark matter-nucleon elastic scattering cross-section for
in the present model is given as
\be
\sigma_{\rm {SI}}\simeq \frac{m_r^2}{\pi} \left (\frac{g_{\bar\chi \chi h}g_{NNh}}{m_h^2}+
\frac{g_{\bar\chi \chi H}g_{NNH}}{m_H^2} \right )^2.
\label{prob28}
\ee
In the above, $m_r$ is the reduced mass $ = \frac{m_{\chi}m_N}{m_{\chi}+m_{N}}$, 
where $m_N$ is the mass of the scattering nucleon (proton or neutron) and
$g_{NNx}$ ($x= h$ or $H$) denotes the effective Higgs nucleon couplings 
expressed as \cite{Tandean}
\be
 g_{NNh}\simeq (1.217k_d^h+0.493k_u^h)\times 10^{-3}\,\, , \hskip 15 pt
 g_{NNH}\simeq (1.217k_d^H+0.493k_u^H)\times 10^{-3}\,\, .
\label{prob29}
\ee
For the case of THDM I, parameters $k_u^h$ and $k_d^h$ in Eq. \ref{prob29} are
given as
\be
k_u^h=k_d^h=\frac{\cos\alpha}{\sin\beta}\,\, ,
k_u^H=k_d^H=\frac{\sin\alpha}{\sin\beta}\,\, .
\label{prob30}
\ee 
and for the case of THDM II these parameters are
\be
k_u^h=\frac{\cos\alpha}{\sin\beta}\,\, ,k_d^h=-\frac{\sin\alpha}{\cos\beta}\,\, ,
k_u^H=\frac{\sin\alpha}{\sin\beta}\,\, ,k_d^H=\frac{\cos\alpha}{\cos\beta}\,\, . 
\label{prob31}
\ee
Using Eqs.~\ref{prob28}-\ref{prob31}, we compute $\sigma_{\rm SI}$ 
for the DM candidate within the framework of our
chosen specific model in this work and compare them with the latest limits for
$\sigma_{\rm SI}$ and $m_{\chi}$ (in $\sigma_{\rm SI}-m_{\chi}$ plane) given by
recent dark matter direct detection experiments namely XENON100 \cite{xenon}
and LUX \cite{lux} \footnote{Both the experimets use liquid Xenon as detection
material.}. In Fig.~\ref{fig2}a-f we plot the variation of DM-nucleon 
scattering cross-section $\sigma_{\rm SI}$ with DM mass ($m_{\chi}$) for the
cases of both type I and type II THDM. The red and blue lines shown in 
Fig.~\ref{fig2}a-f are the limits on DM-nucleon 
cross-section obtained from XENON100 and LUX. The calculations are performed with the
parameter space (such as couplings etc.) of the present model which has already
been constrained by PLANCK results and collider bounds (Fig.~\ref{fig1}a-c). 
Thus the resulting $m_{\chi}-\sigma_{\rm SI}$ parameter space is in agreement
with the bounds from Higgs signal strength ($R_{\rm CMS,ATLAS}$), limits on the 
signal strength on extra Higgs scalar of THDM ($R'_{I,II}\leq0.2$) and also
satisfies DM relic density predicted by PLANCK. Shown in Fig.~\ref{fig2}a and 
Fig.~\ref{fig2}b are the $m_{\chi}-\sigma_{\rm SI}$ parameter space of DM
candidate in type I THDM framework for $m_{H}=$ 150 and 200 GeV respectively.
Needless to mention, parameters used in these two plots are restricted by
$R_{CMS}$, $R'_{I}$ and PLANCK. It is to be noted from Fig.~\ref{fig1}a that
observational results of Higgs signal strength (Fig.~\ref{fig1}a) indicate
that there is no valid parameter space in type I THDM associated with our
fermionic dark matter that corresponds to $R_{\rm ATLAS}$. It is clear from
Fig.~\ref{fig2}a-b that due to the presence of an extra scalar in the model
along with SM Higgs, an extra pole is likely to appear in the mass range 
$m_{\chi}\sim m_{H}/2$ with the normal SM Higgs pole occuring near 
$m_{\chi}\sim m_{h}/2$. This scenario also holds for the case of type II THDM
as well. Study of the plots in Fig.~\ref{fig2}a-b reveals that the fermionic DM
particle $\chi$ in type I THDM can serve as a viable candidate of dark matter
with a sufficient allowed parameter space that is in agreement with latest DM
direct detection experimental results of XENON100 and LUX. Similarly using the
allowed parameter space obtained in Fig.~\ref{fig1}b-c (constrained by DM relic
density, combined Higgs signal strength ($R_{\rm CMS,ATLAS}$) and bound on 
additional Higgs signal ($R'_{II}$)), we plot the viable parameter space in
$m_{\chi}-\sigma_{\rm SI}$ plane for DM in type II THDM (Fig.~\ref{fig2}c-f).
In Fig.~\ref{fig2}c-d, the avialable $m_{\chi}-\sigma_{\rm SI}$ spaces for two
values of the scalar mass $H$, $m_H=$ 150 GeV and 200 GeV respectively are
shown. Each of these plots satisfies the model parameter space constrained by
PLANCK, $R_{\rm CMS}$ and $R'_{II}$. Analogus plots are obtained in 
Fig.~\ref{fig2}e-f but here only $R_{\rm ATLAS}$ is taken into account instead
of $R_{\rm CMS}$. It is obvious from Fig.~\ref{fig2}c-f, that the region of
allowed $m_{\chi}-\sigma_{\rm SI}$ space depends on the mass of the additional
scalar $H$. Fig.~\ref{fig2}c-f also shows that a considerable portion of 
DM-nucleon scattering cross-section $\sigma_{\rm SI}$ of the DM candidate
$\chi$ in type II THDM lies in the allowed region set by XENON100 and LUX 
direct detection experiments. Hence, fermionic dark matter $\chi$ appearing
in type II THDM can be treated as a potential candidate for dark matter.
It is also seen from Figs.~\ref{fig2}a-f that as we do not involve
any condition on DM-Higgs coupling (such as $g_{\bar{\chi}\chi h}=0$ or
$g_{\bar{\chi}\chi H}=0$ \cite{Cai}) for SM like scalar,
the low mass region of dark matter apppering in \cite{Cai} 
($m_{\chi}\leq40$ GeV) is excluded.
\subsection{Indirect search of dark matter : Gamma-ray flux Calculation}
\label{ss:id} 
In indirect detection of dark matter, the experimets look for excess signature
of $\gamma$-ray, neutrino, positron and anti-proton flux that might have
originated from the annihilation of dark matter candidate into  SM particles. 
In this section, we study such excess
$\gamma$-ray flux from the Galactic Centre (GC) region observed by Fermi-LAT
(or FGST) \cite{fgst} assuming that the excess $\gamma$-ray is produced by the
process of dark matter pair annihilation at GC. We consider the particle dark
matter candidate is the fermionic dark matter $\chi$ in the present framework.
In previous works \cite {hooper,hooper1} this $\gamma$-ray is reported to be
in the range 1-10 GeV and it was explained by considering the annihilation
of 10 GeV dark matter at GC.
In order to investigate whether our proposed fermionic DM
candidate $\chi$ in both type I and type II THDM can  account for the observed
$\gamma$-ray excess originating from the inner galaxy 
($5^0$ surrounding the GC), we first  calculate the $\gamma$ flux in inner
galactic region produced from the annihilation of DM candidate $\chi$. We then
add to it the $\gamma$-ray flux arisingout of the known sources present in the
inner galaxy (galactic ridge and point source emission) and compare the 
resultant $\gamma$ flux with the FGST observations of GC gamma-ray flux within
the inner $5^0$ region. The flux for the galactic ridge and point sources of
$\gamma$-ray emission are obtained from Refs. \cite{gr}-\cite{ps2}. The 
differential gamma-ray flux produced  at GC from the annihilation of DM 
particles in a direction that subtends a solid angle $d\Omega$ is given 
as \cite {cirelli}
\begin{table}
\begin{center}
\resizebox{\textwidth}{!}{
\begin{tabular}{|c|c|c|c|c|c|c|c|c|c|c|c|}
\hline
    &     &        &            &                        &                        &              &             &               &                                    & 
                               &                                \\ 
THDM & BP  & $m_H$  & $m_{\chi}$ & $g_{\bar{\chi}\chi h}$ & $g_{\bar{\chi}\chi H}$ & $\sin\alpha$ & $\tan\beta$ & $\sigma_{SI}$ & $\langle \sigma v\rangle_{b\bar b}$&
$\langle \sigma v\rangle_{WW}$ & $\langle \sigma v\rangle_{ZZ}$ \\
      &     & in GeV &   in GeV   &                        &                        &              &             & in cm$^2$     &             cm$^3$/s            &
            cm$^3$/s            &        cm$^3$/s             \\
\hline  
    &  1  &  150.0  &   55.0     &   -4.00e-02    &  -9.00e-02  &   -0.545   &   2.05     &   4.14e-48  & 1.62e-26  & 1.79e-34 & 1.691e-37 \\
  
 I  &     &         &            &                &             &            &            &             &            &           &           \\ 

    &  2  &  200.0  &   90.0     &   2.00e-03     &  1.00e-02   &  -0.399    &   28.63    &   1.08e-48  & 3.43e-29  & 1.10e-26 & 5.52e-27 \\
\hline
    &  3  &  150.0  &   70.0     &   -2.00e-02    &  2.00e-02   &  -0.375    &   2.90     &   2.87e-46  & 1.70e-26  & 3.46e-29 & 5.53e-32 \\
  
 II &     &         &            &                &             &            &            &             &            &           &           \\

    &  4  &  200.0  &   85.0     &   4.00e-02     & -4.00e-02   &  -0.252    &   4.70     &  3.43e-46   & 1.410e-26  & 1.653e-27 & 3.58e-28 \\
\hline
\end{tabular}
}
\end{center}
\caption{Benchmark points (BPs) for type I and type II THDM associated with FDM
used to produce the plots in Fig.~\ref{fig3}a-d. BP1 and BP2 correspond to type
I THDM for $m_{H}=$150 GeV and 200 GeV respectively while BP3 and BP4 signify 
the adopted benchmark points for type II THDM with $m_{H}=$150 GeV and 200 GeV
respectively.}
\label{tab1}
\end{table}

\be
\frac{d\Phi_{\gamma}}{d\Omega dE_{\gamma}}=
\frac{r_\odot \rho^2_{\odot}}{8\pi m_{\chi}^2} J\sum_f {\langle \sigma v \rangle}_f
\frac{dN}{dE_{\gamma}}\,\, ,
\label{prob32}
\ee
where $r_{\odot}$=8.5 kpc is the distance of the Sun from GC and 
$\rho_{\odot}$=0.3 GeV cm$^{-3}$ is the local DM density at solar 
neighbourhood. In Eq.~\ref{prob32},
$\frac{dN}{dE_{\gamma}}$ is the photon spectrum per annhilation of 
DM into final state $f$ with annihilation cross-section 
${\langle \sigma v \rangle}_f$. The astrophysical factor $J$ for annihilating
DM, is given by the relation (with $\rho(r)$) denoting halo density profile)
\be
J=\int_{l.o.s.} \frac{ds}{r_{\odot}} \frac{\rho^2(r(s,\theta))}{\rho^2_{\odot}}\,\, ,
\label{prob33}
\ee 
In the above, the line of sight integral has been performed over an angle
$\theta$. In Eq.~\ref{prob33}, $\theta$ is the angle between the line from earth
to GC and the direction of line of sight at a distance $r$
($r=\sqrt{r_{\odot}^2+s^2-2r_{\odot}s\cos\theta}$) from GC. The gamma-ray flux
originating from an extended region with soild angle $\Delta\Omega$ takes the
form 
\be
\frac{d\Phi_{\gamma}}{dE_{\gamma}}=
\frac{r_\odot \rho^2_{\odot}}{8\pi m_{\chi}^2} \bar{J} \Delta\Omega \sum_f {\langle \sigma v \rangle}
\frac{dN}{dE_{\gamma}}\,\, ,
\label{prob34}
\ee
where $\bar{J}$ is the $J$ factor averaged over a soild angle $\Delta\Omega$
and is expressed as 
\be 
\bar{J}=\frac{2\pi}{\Delta\Omega} \int d\theta \sin\theta J(\theta)\,\, ,
\label{prob35}
\ee
with
\be 
\Delta \Omega =\int_{\theta_{min}}^{\theta_{max}} d\theta \sin\theta\,\, .
\label{prob36}
\ee
In the present work we consider two different DM halo density profiles namely
Navarro-Frenk-White (NFW) \cite{nfw} profile and Einasto (Ein) \cite{ein} 
profile. These density profiles are written in the form
\be 
\rho_{NFW}(r)=\rho_s\frac{r_s}{r}\left(1+\frac{r}{r_s}\right)^{-2}\,\, , \hskip 15 pt
\rho_{Ein}=\rho_s {\rm exp}\left(-\frac{2}{\alpha}\left[\left(\frac{r}{r_s}\right)^{\alpha}-1\right]\right)\,\, . 
\label{prob37}
\ee
\begin{figure}[h!]
\centering
\subfigure[]{
\includegraphics[height=5.5 cm, width=5.5 cm,angle=0]{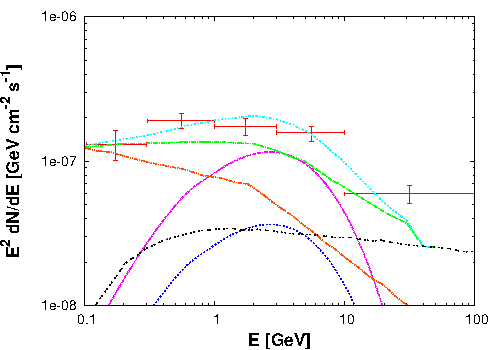}}
\subfigure []{
\includegraphics[height=5.5 cm, width=5.5 cm,angle=0]{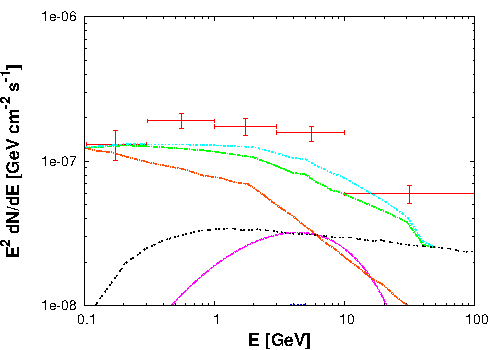}}
\subfigure[]{
\includegraphics[height=5.5 cm, width=5.5 cm,angle=0]{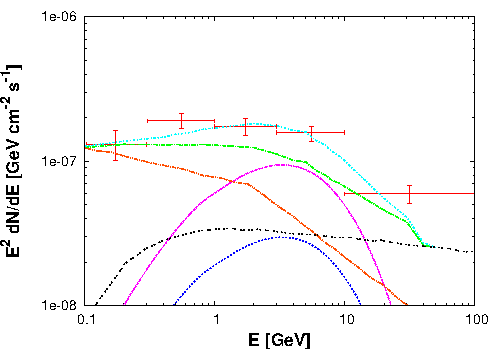}}
\subfigure []{
\includegraphics[height=5.5 cm, width=5.5 cm,angle=0]{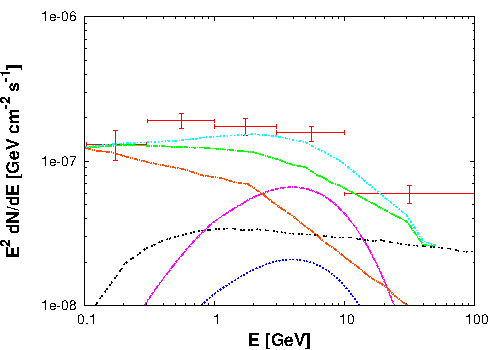}}
\subfigure {
\includegraphics[height=2.5 cm, width=4.5 cm,angle=0]{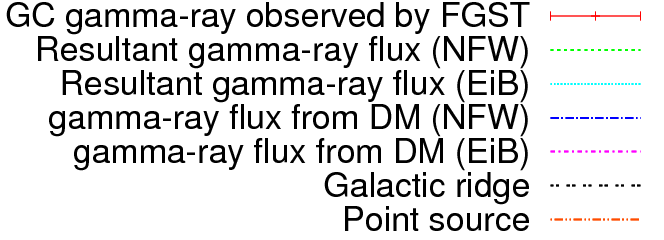}}
\caption{Comparison of the observed gamma-ray flux obtained from FGST with
the resultant gamma-ray flux as generated using the BPs in Table~\ref{tab1}
for the case of fermionic dark matter in type I and type II THDM.}
\label{fig3}
\end{figure}
We have used the numerical values of parameters $r_s$ and $\rho_s$ for the
above halo profiles from 
Ref.~\cite{cirelli}. The chosen values for $\rho_s$in Eq.~\ref{prob37} are 
normalised to produce local DM density $\rho_{\odot}$=0.3 GeV cm$^{-3}$ at 
$r=r_{\odot}$. For the case of Einasto profile, a steeper halo profile with
$\alpha=0.11$ (instead of usual value ($\alpha=0.17$)) is recommended \cite{eib}
when contribution from baryons are also taken into account. This modified
Einasto profile with $\alpha=0.11$ (more genearlly known as EinastoB
or EiB) is chosen in this work. Using Eqs.~\ref{prob32}-\ref{prob37}, we 
calculate the gamma-ray
flux resulting from the annihilation of DM candidate $\chi$ in our model (type
I and type II THDM) from the inner $5^0$ of GC for a chosen set of benchmark
points (BPs) given in Table~\ref{tab1}. We consider two benchmark points for 
each of the type I and type II THDM cases with $m_H=$150 and 200 GeV. These BPs are 
in agreement with the bounds from DM relic density given by PLANCK experiment,
signal strength of SM scalar from LHC, the adopted bound on the signal strength
of non-SM scalar ($R'_{x}\leq~0.2$) and direct detection constraints from LUX.
We now calculate the gamma-ray flux for the benchmark points considered in
Table \ref{tab1} and compare with the observed gamma-ray flux obtained from
FGST data \cite{fgst} \footnote{ It is to be noted that for the allowed 
$m_\chi$-$\sigma_{\rm SI}$ plane shown in Fig.~\ref{fig2}a-f, the scattering 
cross-section $\sigma_{\rm SI}$ could become less than $10^{-48}$ $\rm{cm}^2$
and overlap with the cosmic neutrino scattering region. Direct detection of DM
would be difficult in this region. To avoid this we have chosen the BP's in
Table~\ref{tab1} carefully such that $\sigma_{\rm SI}\ge 10^{-48} \rm{cm}^2$
but consistent with the upper bounds given by XENON100 and LUX}.
The calculated $\gamma$-ray flux for the benchmark 
points considered are plotted in Fig.~\ref{fig3}a-d. As mentioned earlier, the 
calculations are performed for both NFW and EiB dark matter density profiles.
The total $\gamma$-ray flux are then obtained by adding the calculated flux for
dark matter annihilation in GC region (either with NFW or with EiB profile) 
with the galactic ridge and point source data. Therefore, for each plots (a-d)
of Fig.~\ref{fig3} we show two results for total $\gamma$-ray flux (from DM
annihilation + galactic ridge + point sources) that correspond to NFW and EiB
profiles and compare both of them with the Fermi-LAT experimental data points
(with error bars) in Fig.~\ref{fig3}a-d. Also plotted in Fig.~\ref{fig3}a-b the
contribution to the $\gamma$-ray flux from galactic ridge and point sources 
separately. the calculational results shown in Fig.~\ref{fig3}a and 
Fig.~\ref{fig3}b correspond to the first set of benchmark points BP1 and BP2
respectively chosen for type I THDM. Fig.~\ref{fig3}a and Fig.~\ref{fig3}b
shows the results for $m_{H}=$ 150 GeV and 200 GeV respectively. For type
II THDM the set of benchmark points BP3 ($m_{H}=$ 150 GeV) and BP4 ($m_{H}=$
200 GeV) are adopted and the corresponding results are shown in 
Fig.~\ref{fig3}c and Fig.~\ref{fig3}d respectively (Table~\ref{tab1}, 
lower part). Comparison of the plots a and b of Fig.~\ref{fig3} reveals that
in case of our proposed fermionic dark matter in type I THDM scenario, the
experimental data are best satisfied when we choose the non-SM Higgs mass
$m_{H}$ to be 150 GeV and consider Einasto B  (EiB) profile in our
calculations. For the choice of $m_{H}=$ 200 GeV in type I THDM case however the
total flux calculated with either of the NFW or EiB profiles do not agree at
all with the experimental data. Similar plots with BP 3 and BP 4 that 
correspond to type II THDM are shown in Fig.~\ref{fig3}c-d. Plots in 
Fig.~\ref{fig3}c-d show that in the case of our fermionic dark matter in type
II THDM too, total $\gamma$-ray flux generated using EiB profile is in good 
agreement with the observed data when compared to those calculated using NFW
profile. Finally, $\gamma$-ray produced by the annihilation of our proposed
fermionic dark matter in the mass range $\sim$ 50-90 GeV (in type I and type II
THDM) can best explain the experimentally observed $\gamma$-ray excess from GC.
In Table~\ref{tab1} we tabulate the thermally averaged annihilation cross-section
for DM annihilating into $b\bar b$, $W^+W^-$ and $ZZ$ channels along with
model parameters. From BPs in Table~\ref{tab1} one observes that for BP1 and
BP3, DM mainly annihililates into $b\bar b$.
For the case of BP2 (with $m_{\chi}=90$ GeV) the $W^+W^-$ channel
dominates over other annihilation channels thereby reducing the $\gamma$-ray
flux produced for BP2. For BP4 ($m_{\chi}=85$ GeV), DM annihilating into 
$b\bar b$ is dominant although $W^+W^-$ channel also contributes (nearly 10\%
of $\langle \sigma v\rangle_{b\bar b}$). Study of annihilation channels shows
that for DM with smaller masses (BP1 and BP3) primarily tends to annihilate
into $b\bar b$ but with the increase in DM mass other annihilation channels will
open up and contributes significantly.

\begin{figure}[h!]
\centering
\subfigure[]{
\includegraphics[height=5.5 cm, width=5.5 cm,angle=0]{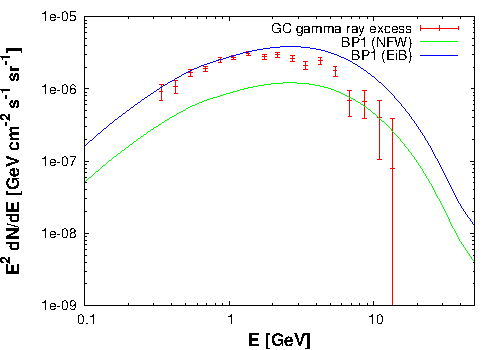}}
\subfigure []{
\includegraphics[height=5.5 cm, width=5.5 cm,angle=0]{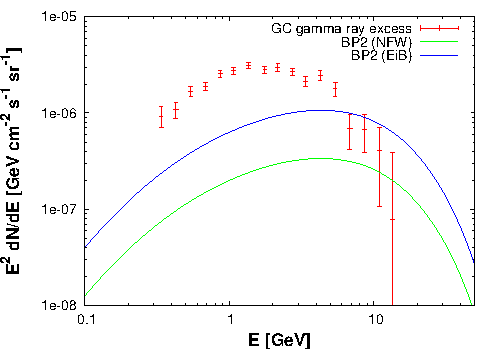}}
\subfigure[]{
\includegraphics[height=5.5 cm, width=5.5 cm,angle=0]{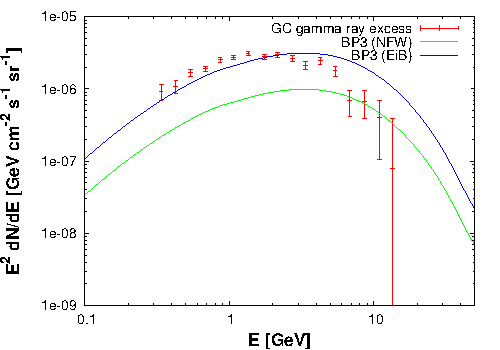}}
\subfigure []{
\includegraphics[height=5.5 cm, width=5.5 cm,angle=0]{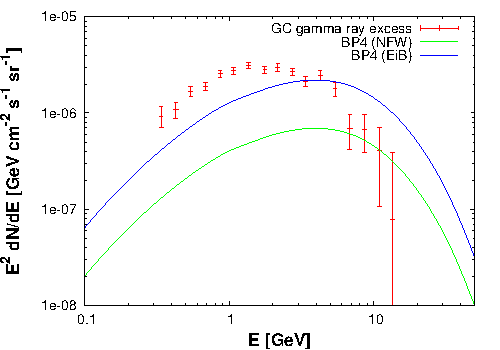}}
\caption{Inner galaxy gamma ray flux obtained using BPs of Table~\ref{tab1}
for fermionic DM in type I and type II THDM compared with the observed
Fermi-Lat data \cite{Daylan:2014rsa}.}
\label{fig4}
\end{figure}
We also like to add that
recent analysis of Fermi-LAT data for gamma rays from inner galaxy has
reported an excess of $\gamma$-rays from Galactic Centre region
\cite{Goodenough:2009gk}-\cite{Daylan:2014rsa} in the energy range 1-3 GeV.
Since no astrophysical phenomena can
explain this excess, the annihilation of DM at the GC region could be responsible
for the same \cite{Daylan:2014rsa}-\cite{Kong:1404}. Many different particle 
physics models for dark matter candidate have been proposed and explored in
order to explain this 1-3 GeV $\gamma$-ray excess 
\cite{Kaye:1310}-\cite{Banik:2014eda}. In this work, we also explore in the
framework of our model, whether the present singlet
fermionic dark matter in type I and type II THDM can 
explain this GC gamma ray excess. We calculate the GC $\gamma$-ray flux 
obtained from the annihilation of our fermionic DM in this present framework.
In doing this we use the benchmark points for the fermionic dark matter 
discussed earlier. The two dark matter halo profiles namely the NFW and EiB 
halo profiles as mentioned above are also used in order to explain the 
1-3 GeV $\gamma$-ray excess from GC. 
Shown in Fig.~\ref{fig4}a-d, the results and their comparisions with data points. 
In the plots of Fig.~\ref{fig4}a-d, the experimental data points are shown in
red whereas the green and blue lines represent the calculational results for
the benchmark points for the cases with NFW and EiB halo profile 
respectively \footnote{All the plots are presented in logarithmic scale. 
Hence only data points with positive flux values are considered.}. 
In Fig.~\ref{fig4}a-b
we plot $\gamma$-ray flux produced due to DM annihilation using BP's adopted
for type I THDM secanrio (i.e., BP1 and BP2) with $m_H=$ 150 GeV and 200 GeV.
From Fig.~\ref{fig4}a-b, it can be observed that for type I THDM the excess of
$\gamma$-ray produced using BP1 is in better agreement with the experimental
results compared to the case with BP2 when EiB profile is considered. Similar
plots for the case of type II THDM are shown in Fig.~\ref{fig4}c-d. In case of
type II THDM, we found that the excess in $\gamma$-ray obtained for BP3 with
EiB halo profile is in good agreement with the experimental results of 
Fermi-LAT in comparision to the results when BP4 is chosen. For both type I and
type II THDM framework with singlet fermion the calculated gamma flux using
NFW profile with BP's are not compatible with the observed
results of Fermi-LAT.          
\section{Discussions and  Conclusions}
\label{S:discussion}
In this work we consider a singlet fermion dark matter in a framework of 
two Higgs doublet model. We have explored the viability of such a fermionic
dark matter in two different types of THDMs namely THDM I and THDM II and
assumed that the new found scalar boson at LHC is one of the two CP even Higgs
occuring in THDM. The fermionic dark matter candidate $\chi$ in our model
couples to the CP even Higgs-scalars appearing in THDM with a 
non-renormalisable dimension five interaction. Hence, DM in the present model
can undergo the process of annihilation into SM particles through Higgs
mediated channels. We solve the Boltzmann equation for the DM candidate $\chi$
to calculate the DM relic density for the case of type I and type II THDM. We
have constrained the model parameter space by PLANCK relic density criterion
for dark matter, bounds on the SM Higgs signal strength obtained from LHC
experiments (CMS and ATLAS) and latest direct detection limits on DM-nucleon
scattering cross-section from XENON100 and LUX results. Since both the models
(type I and type II THDM) involve an extra Higgs boson ($H$), additional bounds
on the signal strength of non-SM scalar due to its non-observance are also
taken into account. Study of the model parameters reveals that an increase in
the mass of $H$ ($m_H$) will result in a decrease in the valid parameter space
for both the THDM's considered. The present analysis indicates that the
fermionic DM $\chi$ in THDM I and II framework (as considered in the work) can
be treated as a possible dark matter candidate satisfying the bounds on DM
relic density, direct detection and Higgs signal stregth results from CMS and
ATLAS. We further test the viability of our model by investigating whether the
DM in present mechanism can produce the observed GC gamma-ray flux predicted by
FGST. We have chosen two sets of benchmark points, each of one for type I and
type II THDM scenarios. We also consider two dark matter halo profiles namely
NFW and Einasto B. We calculate the $\gamma$-ray flux originating from DM
annihilation for these chosen benchmark points. Comparison of the observed FGST
$\gamma$-ray flux with those obtained from the BPs in our model suggests that
the fermionic DM candidate $\chi$ in our framework can better explain the GC
$\gamma$-ray if Einasto B halo profile is considered. The present 
framework of THDM excludes the low mass regime explored in the work \cite{Cai}
when the ad-hoc asumption on the DM-Higgs coupling is relaxed. This also
holds for the case of scalar or vector dark matter candidate explored in the
work \cite{Cai}. 
We have found
that instead of a low mass DM $\sim$ 10 GeV, the fermionic DM in the present 
scenario in the mass range $50~{\rm GeV}\leq m_{\chi}\leq 90~{\rm GeV}$ can also
provide a plausible explanation to the GC $\gamma$-ray observed by Fermi-LAT.    

\end{document}